\documentclass[aps,preprint]{revtex4}

\usepackage{color}
\usepackage{graphicx}

\newcommand{\ba}{\begin{eqnarray}}
\newcommand{\ea}{\end{eqnarray}}

\begin{document}

\date{\today}

\title{Non-Hermitian formalism and nonlinear physics} 

\author{
Hichem Eleuch$^{1,2}$\footnote{email: heleuch@physics.tamu.edu} and 
Ingrid Rotter$^{3}$\footnote{email: rotter@pks.mpg.de; corresponding author}}

\affiliation{
$^1$ Department of Applied Sciences and Mathematics, 
College of Arts and Sciences, Abu Dhabi University, Abu Dhabi, UAE}
\affiliation{$^2$ Institute for Quantum Science and Engineering, Texas A\&M University, College Station, TX 77843,
United States of America}
\affiliation{
$^3$ Max Planck Institute for the Physics of Complex Systems,
D-01187 Dresden, Germany  }

\vspace*{1.5cm}

\begin{abstract}

The non-Hermitian formalism is used at present in many papers for the
description of open quantum systems. A special language developed in
this field of physics which makes it difficult for many physicists to
follow and to understand the corresponding papers. We show that the
characteristic features of the non-Hermitian formalism are nothing but
nonlinearities that may appear in the equations when the Hamiltonian
is non-Hermitian. They are related directly to singular points (called
mostly exceptional points, EPs). At low level density, they may cause
counterintuitive physical results which allow us to explain some
puzzling experimental results. At
high level density, they determine the dynamics of the system.

\end{abstract}

\maketitle

\vspace{1cm}

\section{Introduction}
\label{intr}

The properties of non-Hermitian operators are studied today in many
papers. They appear originally in the description of open quantum
systems such as mesoscopic devises, which assumed an unprecedented
interdisciplinary character. These systems may be defined as
consisting of some localized microscopic  region that is embedded in
an external (natural) environment which is independent of the
system. This environment always exists and can never be deleted. It is
extended in contrast to the system which is localized.
	
It is obvious that such a system can be described successfully by
means of a projection operator technique: one of the operators
projects onto the localized system while another one projects onto the
extended environment. Both parts are described, as usual, separately
by Hermitian operators. The influence of the other subspace onto the
considered one causes, however, the Hamiltonian of the considered
subsystem to be non-Hermitian. For details see  \cite{top,ropp}.
Meanwhile, a special language developed in the non-Hermitian quantum physics
due to which it is difficult
for non-specialists to read and to understand the contents of these papers.    

The reason for the special language is objective. It arises from the
mathematical problems which are inherent in the formalism of
non-Hermitian operators and are unknown and counterintuitive in the
formalism of Hermitian operators. 
Details can be found in e.g. the reviews
\cite{top,ropp}. Nevertheless, the special features
of the notions used in non-Hermitian physics are not unique. They can
be related to well-known phenomena as will be shown in the following.

The only difference between a Hermitian operator $H$ and a
non-Hermitian operator
$\cal H$, which is usually known, is that the eigenvalues $E_i$ of $H$
are real while the eigenvalues ${\cal E}_i$  of $\cal H$  are usually
complex, and only sometimes real. The last case (real eigenvalues) is
known to appear in calculations for concrete realistic systems. The
corresponding eigenstates  are called
usually {\it bound states in the continuum} \cite{wintgen1,wintgen2}. 

The possible coalescence of two eigenvalues of $\cal H$ 
that may appear at a certain value of the considered parameter,
is almost unknown  although it is exactly this
property of  $\cal H$ that determines the interesting  properties of
non-Hermitian physics which are not only unexpected but mostly
counterintuitive. They allow us to explain different
experimental results which remained puzzling in Hermitian quantum
physics \cite{top,ropp,entropy}. 
According to the obtained results, the nature of the
parameter may be very different.

The problem that two
eigenvalues of a non-Hermitian operator $\cal H$ may coalesce 
is known in mathematics since many years \cite{kato}. The
coalescence occurs formally at one point,
which is called mostly exceptional point (EP). In physical systems, 
it is however not enough to know the properties of the
system at one point. Of interest are rather the properties of the
system in a certain finite range of the considered parameter. 
The influence of the EP can be seen, indeed,
in a finite neighborhood around this point as the results of
many numerical studies have shown \cite{top}.
 
The coalescence of two
eigenvalues ${\cal E}_{1,2}$ is different from a possible degeneration
of two eigenvalues $E_{1,2}$
of a Hermitian operator $H$. While the two corresponding
eigenfunctions $\phi_{1,2}$ of $H$ are always different from one
another, the two eigenfunctions  $\Phi_{1,2}$
of $\cal H $ are related to one another by the special relation
\begin{eqnarray}
\label{wf1}
\Phi_1 \to \pm~i~\Phi_2  
\end{eqnarray}
which holds true in the very neighborhood of the EP where the two
states coalesce.  
A consequence of the relation (\ref{wf1}) is that the phases of the
eigenfunctions of $\cal H$  are, generally, not rigid. 
The wavefunctions of two eigenstates of $\cal H$ are almost
orthogonal at large distance (in energy) from one 
another, and related to one another by (\ref{wf1}) in approaching
an EP \cite{top}.
This means that close to the EP the two eigenfunctions are proportional. 
In other words, the eigenfunctions $\Phi_i$ behave according to 
Jordan forms, for details see \cite{gurosa}.
 
We will show in the following that
most of the new terms introduced and used
in the formalism of non-Hermitian operators 
can be replaced by the term {\it nonlinearity}.
The best example is the basic concept {\it exceptional point, EP,} itself. 
At the parameter value at which the EP appears, 
the eigenvalue trajectories of the two states
(traced as function of the considered parameter)
reverse their orientation
and do not cross \cite{top}. Here, 
\begin{eqnarray}
\label{wf2}
{\cal E}_1 \to ~\pm ~i~ {\cal E}_2
\end{eqnarray}
holds true for the (complex) eigenvalues ${\cal{E}}_i$ of $\cal H$,
what corresponds to (\ref{wf1}). Such a
parametrical change of the two eigenvalue trajectories 
is impossible if the movement occurs linearly as a
function of the considered parameter. The change 
is possible only by means of a nonlinear term that is involved in the 
equations describing the trajectories of the eigenvalues
as a function of this parameter.
As usual, this nonlinearity in the 
parametrical dependence of the eigenvalue trajectories influences 
the behavior of the trajectories not only at a certain point 
(at the EP) but also that in the not too far vicinity.  

This fact shows that it is even better to use the expression
{\it nonlinearity} instead of the notion {\it point (EP)}. 
According to the usual understanding,
a nonlinearity influences the behavior of the system in a finite
parameter range.
The notion EP suggests however that all the changes in the behavior
of the system appear only at one point. This will,
of course, never take place in a realistic physical system.

It should be mentioned here that the notation EP is even somewhat
misleading because a single EP has nothing in common with a critical
point at which the properties of the system change. The two eigenvalue
trajectories move independently of one another beyond the influence of
the EP. Changes of the system may appear only under the influence of
two and more EPs.
 
The paper is organized in the following manner.
In Sect. \ref{desc}, the non-Hermitian Hamiltonians used in the 
most commonly used descriptions of quantum systems are given. One of them 
is a genuine
non-Hermitian operator which allows us to study in detail characteristic 
features of the solutions. Another one is an effective Hamiltonian which
is suitable, and often used, in numerical calculations on realistic systems. 
Additionally, we mention the nonlinear formalism used at high level density.
In the following Sect. \ref{traj}, we show numerical results obtained from a 
calculation with a varying number of states in a certain finite energy window. 
The results of these calculations illustrate the influence of resonance 
overlapping onto the results and the role of nonlinearities in the basic 
equations.   The results are summarized and discussed in the last section 
\ref{concl}.

\section{Description of realistic  open quantum systems}
\label{desc}

\subsection{Genuine non-Hermitian operator}
\label{gen}

A genuine description of an open quantum system with two eigenstates
of $\cal H$ starts \cite{proj10} from
\begin{eqnarray}
\label{ham2}
{\cal H}^{(2)} = 
\left( \begin{array}{cc}
\varepsilon_{1} \equiv e_1 + \frac{i}{2} \gamma_1  & ~~~~\omega   \\
\omega & ~~~~\varepsilon_{2} \equiv e_2 + \frac{i}{2} \gamma_2   \\
\end{array} \right) \; .
\end{eqnarray}
Here, $\varepsilon_i$ are the complex eigenvalues of the basic 
non-Hermitian operator $\cal H$ \cite{comment1}.
The $\omega$ stand for the coupling matrix elements of the two
states via the common environment. Their mathematical
expression is derived  in Sect. 3 of \cite{top}. They
are complex where Re($\omega$) is the principal value integral 
and Im($\omega$) is the residuum \cite{top}.
The imaginary part is responsible for coherent processes 
occurring in the system, while the real part contains
decoherences. In (\ref{ham2}), the two nondiagonal matrix elements
$\omega$ are equal since we consider, in the present paper,  
the description of an open quantum system. The 
corresponding expression $\cal H$ for this case is derived in \cite{proj10}.

The eigenvalues ${\cal E}_i \equiv E_i + \frac{1}{2} \Gamma_i$ 
of ${\cal H}^{(2)}$ are, generally, complex:
\begin{eqnarray}
\label{eig1}
{\cal E}_{1,2} \equiv E_{1,2} + \frac{i}{2} \Gamma_{1,2} = 
\frac{\varepsilon_1 + \varepsilon_2}{2} \pm Z 
\end{eqnarray}
with
\begin{eqnarray}
\label{eig2}
Z \equiv \frac{1}{2} \sqrt{(\varepsilon_1 - \varepsilon_2)^2 + 4
  \omega^2}
\; .
\end{eqnarray}
Here,  $E_i$ is the energy and $\Gamma_i$  the width of the eigenstate $i$. 
With $Z\to 0$, the system approaches an EP.
Here, the eigenvalues ${\cal E}_i $ remain complex, generally,
${\cal E}_{1,2} \to (\varepsilon_1 + \varepsilon_2)/2$
according to (\ref{eig1}). Signatures of the EP can
be seen in the vicinity around $Z=0$. 

A Hamiltonian of the type (\ref{ham2}) can be used, in principle, for 
numerical calculations also for a system when the
number of levels with
the complex energies $ \varepsilon_{i} \equiv e_i + \frac{i}{2}
\gamma_i $ is larger than 2. These calculations are however 
complicated and costly. At high level density, 
nonlinearities appear which are caused by several EPs and overlap. 
For numerical examples see Sect. \ref{traj}.

\subsection{Effective Hamiltonian}
\label{eff}

It is more convenient to perform the numerical calculations for a
realistic open quantum system by means of the so-called effective
Hamiltonian \cite{top}
\begin{eqnarray}
H_{\rm eff} = H_B + \sum_C V_{BC} \frac{1}{E^+ - H_C} V_{CB} \, . 
\label{heff}
\end{eqnarray}  
Here, $H_B$ is the Hamilton operator describing the closed 
(isolated) system with discrete states,
\ba
\label{hb}
(H_B - E_i^B)~\Phi_i^B =0 \, ,
\ea
$G_P^{(+)} \equiv (E^+ - H_C)^{-1}  $ is the Green function in the continuum
(environment)
and $V_{BC},~V_{CB}$ describe the coupling of the closed system to the 
continuum. Further, $H_C$ is the Hamiltonian describing the
environment of decay channels. The wave functions $\xi_E^c$ of the 
scattering states (channel wave functions) follow from
\ba
\label{hc}
(H_C-E)~\xi_E^c =0 \, .
\ea
The solutions of (\ref{hb}) are orthonormalized according to the Kronecker
delta $\delta_{ik}$  and those of (\ref{hc}) according to the
Dirac $\delta$ function $\delta(E-E')$ in each channel $c$. 
The solutions of (\ref{hb}) are, of course, nothing but
the solutions of the unperturbed Hamiltonian of the closed system.
$H_C$ is the Hamiltonian of the environment and does not include any
contributions of the (unperturbed) Hamiltonian $H_B$ of the system. 

The advantage of using 
the effective Hamiltonian (\ref{heff}) for numerical calculations is,
above all,
that the results obtained for the corresponding closed system are a part of
the solution.   
The non-Hermitian (second) part of (\ref{heff}) appears as a
correction to the results of a standard calculation. It is caused by the 
embedding of the system into an environment. The
expression (\ref{heff}) is used  in the
numerical calculations performed 
already many years ago for concrete systems.  
Meanwhile numerical results
are obtained for very different systems and by different authors.
For references see \cite{top} and also \cite{ropp}. 
We underline however once more that $H_{\rm eff}$
  describes the system
 approximately in contrast to the genuine operator (\ref{ham2}) which 
represents the exact Hamiltonian.

The properties of the eigenvalues and eigenfunctions of $\cal H$
which are found in the numerical calculations based on (\ref{heff}), are
sketched in Sect. \ref{intr}. Also the meaning of the EPs is discussed
in this introductive section.   

The limiting case of vanishing coupling between system and environment
is described in these calculations  
as a transition of the open system to a closed system, see
the basic equation (\ref{heff}). The formalism does therefore not
consistently describe the system 
as an open quantum system (see the corresponding theoretical and
experimental results obtained for a real existing system with very
small coupling to the environment \cite{savin1,savin2}).
It is however very convenient for the description of realistic open
quantum systems at low level density. Here it provides  reliable results.

\subsection{Nonlinear Schr\"odinger equation}
\label{nonlin}

As mentioned in the Introduction, nonlinear terms appear in the
Schr\"odinger equation in a finite neighborhood of EPs. 
Nevertheless, the results obtained numerically
for open quantum systems, are very seldom discussed and
interpreted by means of
nonlinearities in the basic equations of
the non-Hermitian formalism. The reason is the following.

At low level
density, the different nonlinear parameter ranges are well separated
from one another, and the system is described well and clearly by 
the effective Hamiltonian $H_{\rm eff}$, eq. (\ref{heff}). This
Hamiltonian contains the relevant nonlinear terms at certain parameter
values. 

Only at high level density, the different ranges of the influence of
nonlinear terms overlap. In such a case, the system is described, 
of course, best
by a nonlinear Schr\"odinger equation in which the
nonlinear terms are involved in an averaged manner (in the considered
parameter range).

\section{Eigenvalue trajectories at different level density}
\label{traj}

For illustration, we show in this section some numerical results
obtained by using the genuine non-Hermitian operator of the type
(\ref{ham2}).
The different calculations are performed with a different number of
states all of which are located in the same fixed energy window. In
any case, the individual resonances overlap.
First, the number of states is three. Then, the number of states is
increased up to six by adding one state at any one time.
By this means, the degree of overlapping of the states is varied and
its influence on the trajectories of eigenvalues and eigenfunctions of
$\cal H$ can be studied.
The Hamiltonian (including the coupling term $\omega$) is
  the same in the different calculations.

\begin{figure}[ht]
\vspace{-1cm}
\begin{center}
\includegraphics[width=13.cm,height=14.cm]{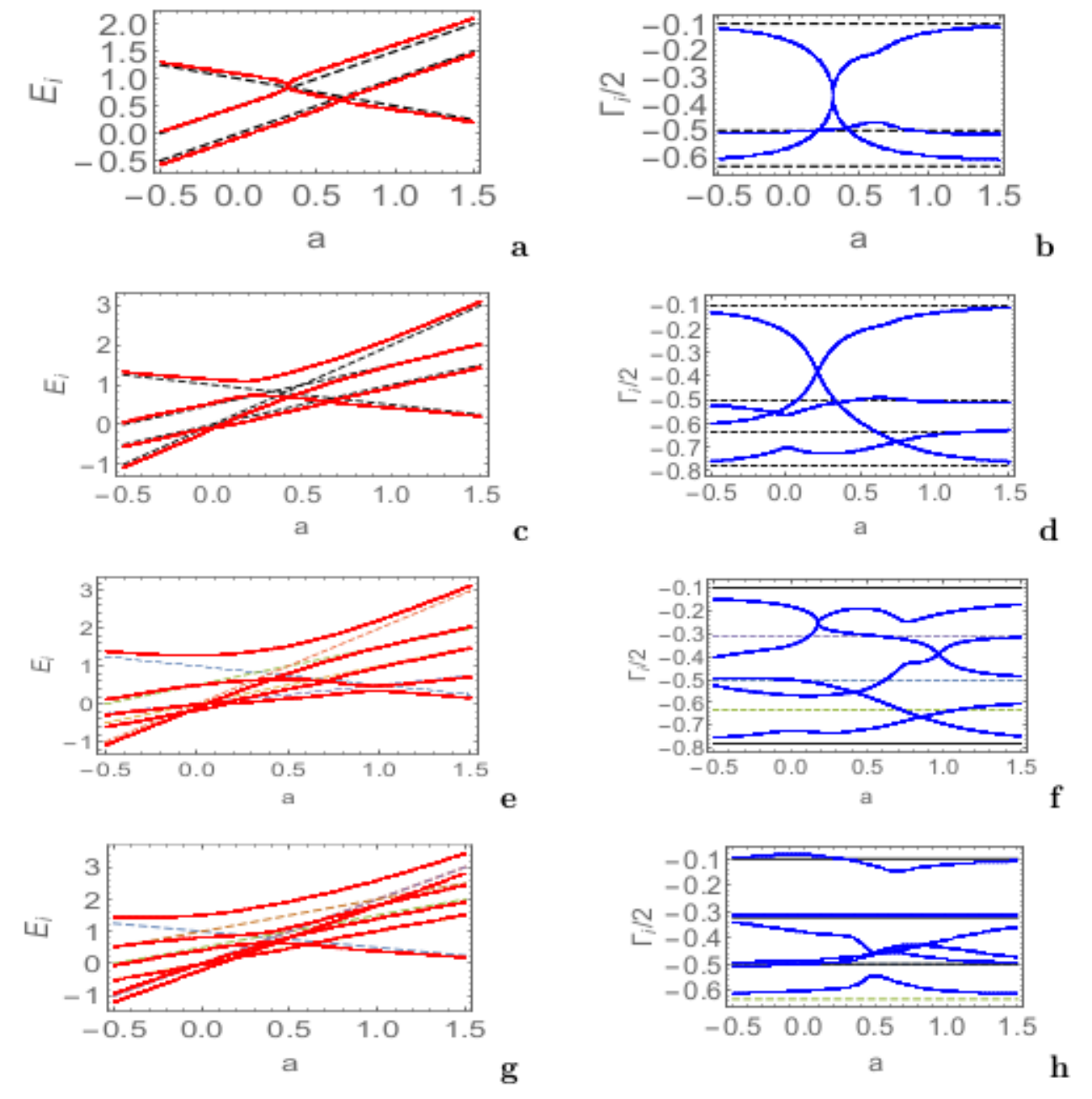} 
\end{center}
\caption{\small The eigenvalue trajectories ${\cal E}_i \equiv E_i + i/2
 \Gamma_i$ of a different number $N$ of states located in
a certain energy window, $N= ~3, ~4, ~5$ and 6.  In all cases, the
individual 
resonance states overlap. The parameters of the plots are:
 $e_1= 1 - a/2 $; $e_2= a$; $e_3=1/2 + a$;
$e_4= 2a$; $e_5= 2a$; $e_6= a + 1$;
$\gamma_{1}/2 = -.5$; $\gamma_{2}/2 = -0.1$;
$\gamma_{3}/2 = -0.6335$; $\gamma_{4}/2 = -0.323$;
$\gamma_{5}/2 = -0.31$; $\gamma_{6}/2 = -0.5$;
The off-diagonal elements of the coupling matrix are $\omega_{ij}=0.2$.}
\label{fig1}
\end{figure}

\begin{figure}[ht]
\vspace{-1cm}
\begin{center}
\includegraphics[width=5.5cm,height=14cm]{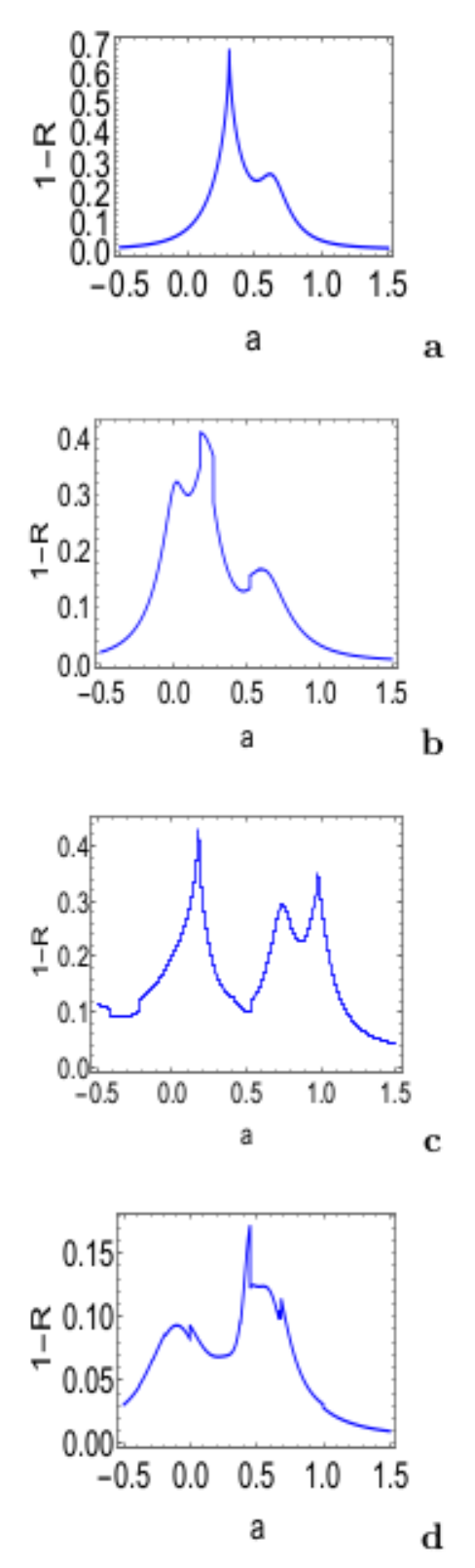}
\end{center}
\caption{\small The eigenfunction trajectories
corresponding to the eigenvalue trajectories of Fig. \ref{fig1}.
The $R$ are averaged phase rigidities obtained by averaging over the
individual phase rigidities $r_i$ of neighboring
states in the considered energy interval.
The 1-R averaged over the considered interval are 
 ~0.119819 for $N=3$;  ~0.122872 for $N=4$;  ~0.164372 for $N=5$; 
 ~0.0634185 for $N=6$.
}
\label{fig2}
\end{figure}

In Fig. \ref{fig1}, the eigenvalue trajectories obtained from a
calculation with respectively $N=3, ~4, ~5$
and $6$ states are shown. The differences of the results arise from
the different degree of overlapping of the states in the different
subfigures.
The corresponding eigenfunction trajectories are shown  in Fig. \ref{fig2}.
The averaged phase rigidities $R$ are the individual phase rigidities 
$r_i$ of the states $i$ averaged over the number of neighboring states
in the considered energy interval.

On the one hand, 
the value $R$ is  an indicator for the difference between the
description of the system by a non-Hermitian Hamiltonian 
and that by a  Hermitian one. In other words
when the value of $R$ is near to 1, the
difference between the description of an open quantum system by a
Hermitian  and that by a non-Hermitian Hamiltonian is not remarkable. 
According to our results, 
this happens only when the number $N$ of states is really large, see
Fig. \ref{fig2}.d. 

On the other hand, the value $R$ contains also information on the role of 
nonlinear terms which are involved in the equations. As expected,
$R$ decreases first with the number $N$ of states in the considered 
parameter region, see the caption of Fig. \ref{fig2}. For $N\to 6$,
however, $R\to 1$ according to Fig. \ref{fig2}.d. 

The results show a similar behavior of all eigenvalue and
eigenfunction trajectories in the considered energy window.  It is the
behavior characteristic of interferences, and the EPs cannot be
identified at large $N$.
Here, hints to their existence even vanish.
The interferences are caused by the individual
contributions from all the resonance states at each point in the
considered energy window. That means the
individual resonance states contribute to the cross section not only
in a restricted small parameter range near to an EP.
They contribute rather to the cross section in a relatively large
parameter range around the EPs,
and the contributions of the different resonance states overlap (when
the level density is sufficiently large). Such a behavior is, of
course, described best by means of the different nonlinear
contributions (caused by the individual EPs) to the cross section.
     
Most interesting are the results for $N=6$. Additionally to the
results shown in Figs. \ref{fig1} and \ref{fig2}, we performed
calculations also for resonance states different from those in these
figures, such that the results are of broader meaning. In all cases,
the features typical of the eigenfunctions of a non-Hermitian operator
vanish at large $N$. The phases of the eigenfunctions become rigid in
the energy window considered, i.e. they behave like the phases of the
eigenfunctions of a Hermitian operator. Correspondingly, the averaged
value $R$ approaches the value $1$. Obviously, the system is described
at high level density best by the Hermitian formalism with inclusion
of nonlinear effects.

\section{Conclusions}
\label{concl}

The numerical results represented in the two figures show very clearly
that the nonlinearities arising from the EPs determine the behavior
of realistic  physical systems at high level density. It is therefore
justified to replace the {\it concept EP} (defined at a certain point in the
parameter space) by the
{\it synonym nonlinearity} (defined in a much broader parameter range).
Such a replacement corresponds to the real situation appearing in 
(small) open quantum systems at high level density.


\begin{thebibliography}{99}
\bibitem{comment1} In contrast to the definition that is used in, 
for example, nuclear physics, we define the complex energies 
before and after diagonalization of ${\cal H}$ by 
$\varepsilon_k = e_k + \frac{i}{2} \gamma_k$ and  
${\cal E}_k = E_k + \frac{i}{2} \Gamma_k$, respectively, 
with $\gamma_k \le 0$ and $\Gamma_k \le 0$ for decaying states. 
This definition will be useful when discussing systems with  gain
(positive widths) and  loss (negative widths).
\end{thebibliography}
\end{document}